\documentstyle[aps,12pt]{revtex}

\newcommand{\Journal}[4]{\textit{#1}\ \textbf{#2}, #3 (#4)}

\newcommand{\JHEP}[3]{\Journal{JHEP}{#1}{#2}{#3}}

\newcommand{\NPB}[3]{\Journal{Nucl.\ Phys.}{B#1}{#2}{#3}}

\newcommand{\PLB}[3]{\Journal{Phys.\ Lett.}{#1B}{#2}{#3}}

\newcommand{\PRD}[3]{\Journal{Phys.\ Rev.}{D#1}{#2}{#3}}

\newcommand{\PRL}[3]{\Journal{Phys.\ Rev.\ Lett.}{#1}{#2}{#3}}

\newcommand{\hepph}[1]{{hep-ph/#1}}
\newcommand{\hepth}[1]{{hep-th/#1}}

\begin{document}

\preprint{ UW-PT/01-04}

\begin{center}
\Large\bf
Wave Function of the Radion in the dS and AdS Brane Worlds 
\end{center}

\author{Z. Chacko \footnote{zchacko@phys.washington.edu} and \ Patrick
J. Fox \footnote{pjfox@phys.washington.edu}}

\smallskip
\address{Department of Physics, Box 351560\\
University of Washington\\
Seattle, Washington, 98195, USA}

\begin{abstract} 

We study the linearized metric perturbation corresponding to the radion
for the generalization of the five dimensional two brane setup of Randall
and Sundrum to the case when the curvature of each brane is locally
constant but non-zero.  We find the wave fuction of the radion in a
coordinate system where each brane is sitting at a fixed value of the
extra coordinate.  We find that the radion now has a mass$^2$, which is
negative for the case of de Sitter branes but positive for anti de Sitter
branes. We also determine the couplings of the radion to matter on the
branes, and construct the four dimensional effective theory for the radion
valid at low energies. In particular we find that in AdS space the wave
function of the radion is always normalizable and hence its effects,
though small, remain finite at arbitrarily large brane separations.

\end{abstract}



\section{Introduction}

In the last three years there has been considerable interest in theories
where the standard model fields are localised to a brane in a higher
dimensional space, motivated by the work of Arkani-Hamed, Dimopoulos and
Dvali \cite{ADD} and also Randall and Sundrum \cite{RS1}\cite{RS2}. In
particular, Randall and Sundrum presented a five dimensional model with
two branes in which the hierarchy problem can be solved due to the
exponentially changing metric along the extra dimension. We will refer to
this model as RS1 \cite{RS1}. In this model the tensions of both branes
have to be fine tuned with respect to the bulk cosmological constant in
order to get solutions which have the Minkowski metric along the four non
compact dimensions. The spacing between the branes is also undetermined
and corresponds to a massless scalar field, called the radion, in the four
dimensional effective theory. Randall and Sundrum also investigated a
model with a single brane and an infinitely large extra dimension, which
we will refer to as RS2 \cite{RS2}.

In RS1 if the brane tensions are not fine tuned with respect to the bulk
cosmological constant then the solutions obtained have either the de
Sitter or the anti de Sitter metric along the four non compact dimensions
\cite{Kaloper}. The branes are therefore bent and the brane spacing is now
fixed in terms of the parameters of the theory. Solutions of this type
with AdS metric along the four noncompact dimensions can have a minimum or
`bounce' of the warp factor with the consequence that four dimensional
gravity arises from the exchange of both massive and massless modes \cite{KR}, 
\cite{Kogan2}. In earlier models with Minkowski metric along the
four noncompact dimensions that exhibit such properties
\cite{Kogan1}, \cite{GRS} the radion plays a crucial role \cite{PRZ} in cancelling the
unwanted polarizations of the 5-D graviton. However it has a negative
kinetic term in the 4-D effective theory and is therefore a ghost
\cite{PRZ}, see also \cite{CEH}, \cite{DGM}.  However the physics of the
radion in these AdS models with a bounce in the warp factor has not been
investigated.\footnote {For a discussion of the graviton polarization properties
  in AdS space see \cite{P}.}

In this paper we study the linearized metric perturbation corresponding to
the radion in bent brane models and determine its mass and and its
couplings to physics on the branes. We show that for the case of 4D de
Sitter space along the noncompact dimensions the radion has negative
mass$^2$ in agreement with an earlier result \cite{GS}, while for 4D anti
de Sitter it has a positive mass$^2$.  This brane configuration is
therefore unstable in de Sitter space. We further construct the low energy
4D effective theory for the radion in these spaces. From the effective
theory we find that for AdS models with a bounce in the warp factor the kinetic
term for the
radion is positive and hence it is not a ghost. Surprisingly, the
effective theory shows that the radion does not completely decouple from
matter on the visible brane even when the branes are widely spaced and
separated by a bounce in the warp factor. This is because in AdS space the
radion wave function is normalizable at arbitrarily large brane
separations.

In the Randall Sundrum model the radion couplings have been investigated
by Charmousis et al. \cite{CGR} in a coordinate system where both the
branes sit at fixed values of the extra coordinate. Similar coordinate
systems have been employed by other authors for considering more
complicated models \cite{CGK}, \cite{PRZ}. It is our opinion that the
physics of the radion is particularly clear in such a basis, and so we
shall find and employ such a coordinate system for studying the bent brane
model. For an interesting alternative approach which directly yields the
effective theory of the radion in the Randall Sundrum model see \cite{LS}, 
\cite{Goldberger}, \cite{Csaki:2000mp}.

\section{The Radion Wave Function}

The action for our system is 

\begin{equation}
S = \int d^4 x d{r} \left[\sqrt{-G} (2 M^{3} R - \Lambda_B) 
-\delta(r)  \sqrt{-\bar{G}} \bar{\Lambda}_0 - 
\delta(r - a)  \sqrt{-\bar{G}} \bar{\Lambda}_{A}\right] 
\end{equation}

In our notation we label a general coordinate by $x^M$ where $M$ runs from
0 to 3 and 5. Our usual four dimensional coordinates are represented by
$x^\mu$
while the extra coordinate $x^5 = r$ is compact and runs from $-a$ to $a$.
Furthermore we make the identification of $(x,r)$ with $(x,-r)$ so that we
are working in the space $S^1$/{\bf $Z_2$}. The branes are located at the
orbifold fixed points $r = 0$ and $r=a$. The components of the
5-dimensional metric tensor are in general represented by $G_{MN}$ while
the induced metric on each brane is represented by ${\bar G}_{\mu
\nu}$. The
Planck scale of the higher dimensional theory is represented by $M$, the
bulk cosmological constant by $\Lambda_B$ and the wall cosmological
constants by $\bar{\Lambda}_0$ and $\bar{\Lambda}_A$.

It has been shown \cite{Kaloper} that this system admits solutions with
constant brane curvature for general values of the brane tensions. These
solutions preserve either four dimensional de Sitter, anti de Sitter or
Poincare invariance along the four non-compact dimensions.

To find these solutions we parametrize the metric by 

\begin{equation}
ds^2 = f(r) g_{\mu \nu} dx^{\mu} dx^{\nu} + dr^2
\end{equation}

where 

\begin{eqnarray}
g_{\mu \nu} dx^{\mu} dx^{\nu} &=& -dt^2 + e^{2Ht}(dx^2 + dy^2 + dz^2)  
\qquad {\rm de \;
Sitter} \\
g_{\mu \nu} dx^{\mu} dx^{\nu} &=& -dt^2 + dx^2 + dy^2 + dz^2  
\qquad \qquad \; \; {\rm Poincare} \\ 
g_{\mu \nu} dx^{\mu} dx^{\nu} &=& dx^2 + e^{2Hx}(-dt^2 + dy^2 + dz^2)  
\qquad {\rm anti \; de \; Sitter}        
\end{eqnarray}
 
On substituting this form into the Einstein equations in the bulk

\begin{equation}
2 M^{3} \left( R_{MN} - \frac{1}{2} g_{MN}R \right) = -\frac{1}{2}
g_{MN} \Lambda_B\ 
\end{equation}

we get for the de Sitter case

\begin{eqnarray}
\frac{3}{2} \frac{f''}{f} - 3 \frac{ H^2}{f} &=& \frac{3}{2} \alpha^2 \\
\frac{3}{2} \left(\frac{f'}{f}\right)^2 - 6 \frac
{ H^2}{f}  &=&
\frac{3}{2} \alpha^2
\end{eqnarray}

The analogous equations for the anti de Sitter and Poincare invariant cases
may be obtained by setting $H^2 \rightarrow -H^2$ and $H^2 \rightarrow 0$
respectively. Here we have defined $\alpha$ by

\begin{equation}
\frac{3}{2} \alpha^2 = -\frac{\Lambda_B}{4M^{3}}
\end{equation}

The solutions of these equations are 

\begin{eqnarray}
f =& 4 \frac{H^2}{\alpha^2} {\rm sinh^2} 
\left[\frac{1}{2}\alpha(r-r_0)\right]
& \qquad {\rm de \; Sitter} \\
f =& e^{ \alpha r} \; \; {\rm or} \; \; e^{ -\alpha r} & \qquad {\rm
Poincare}\\
f =& 4\frac{H^2}{\alpha^2} {\rm cosh^2} 
\left[\frac{1}{2}\alpha(r-r_0)\right]
& \qquad {\rm anti \; \rm de \; Sitter}
\end{eqnarray}

Here $H, r_0$ and the brane spacing a are determined by the jump condition
at each brane, and the requirement that $f$ be normalized to 1 at the
position of the visible brane. Since there are a total of three conditions
to be met and three parameters which can be adjusted to satisfy them no
fine tuning is required.

We expect the radion mode to correspond to a fluctuation about these
solutions that alters the positions of the branes and which transforms
like a scalar in the four dimensional subspace.

Hence we investigate perturbations of the form

\begin{equation}
ds^2 = \left[f(r) + \epsilon (r) \psi (x) \right] g_{\mu \nu} dx^{\mu}
dx^{\nu} + \left[ 1 + \lambda (r) \psi(x) \right] dr^2
\end{equation}
  
This form guarantees that $\psi (x)$ will transform as a scalar on the
brane. It is this field that we expect to play the role of the radion. We
substitute this into the linearized Einstein equations in the bulk, 
which are

\begin{equation}
\delta R_{MN} = - \alpha^2 \delta G_{MN} 
\end{equation}

Now the $\delta R_{\mu \nu}$ equation will contain a term of the form

\begin{equation}
\delta R_{\mu \nu} = - \left[\frac{\epsilon}{f} + \frac{1}{2}\lambda
\right] \nabla_{\mu}   
\nabla_{\nu} \psi  + \; . \; . \; .
\end{equation}

Since all the other terms in this equation as well as the source itself
are proportional to $g_{\mu \nu}$ we look for a consistent solution for
which this term vanishes. Hence we proceed by setting

\begin{equation}
\label{eq:f}
\frac{\epsilon}{f} + \frac{1}{2} \lambda = 0
\end{equation}

Then the Einstein equations for the de Sitter case take the form 

\begin{eqnarray}
\mu \nu &:& \; - \frac{1}{2} \left(\frac{\epsilon}{f} \right)''
-2 \frac{f'}{f} \left(\frac{\epsilon}{f} \right)' +
\left[\lambda - \frac{\epsilon}{f} \right] \frac{3 H^2}{f} + \lambda
\alpha^2 - \frac{m^2}{2f}\frac{\epsilon}{f} + \frac{1}{4} \frac{f'}{f}
\lambda' = 0 \\
\mu 5 &:& \; \frac{3}{2}\left(\frac{\epsilon}{f} \right)' - 
\frac{3}{4} \lambda \frac{f'}{f} = 0 \\
55 &:& \; -2 \left(\frac{\epsilon}{f} \right)'' - 2
\left(\frac{\epsilon}{f} \right)' \frac{f'}{f} + 
\lambda' \frac{f'}{f} - \lambda \frac{m^2}{2f} + \alpha^2 \lambda = 0
\end{eqnarray}

As before, the corresponding equations for the anti de Sitter and
Minkowski cases may be obtained by setting $H^2 \rightarrow -H^2$ and $H^2
\rightarrow 0$ respectively. The quantity $m^2$ is defined by $g^{\mu \nu}
\nabla_{\mu} \nabla_{\nu} \psi = m^2 \psi$ in the above equations.

Since the $R_{\mu 5}$ equation is first order it can easily be solved
using (\ref{eq:f}). The solution, up to an overall multiplicative
constant,
is simply

\begin{eqnarray}
\label{eq:solution}
\epsilon &=& 1 \\
\lambda  &=& - \left(\frac{2}{f} \right)
\end{eqnarray} 

Notice that neither the $R_{\mu 5}$ equation or equation (\ref{eq:f}) 
involves $H^2$. This means that the solution in this form is in fact
valid for Minkowski and anti de Sitter 3 branes as well. It is 
straightforward to verify that this is in perfect agreement with the
result of Charmousis et al. \cite{CGR} for the case of Minkowski 
branes.

The solution above satisfies the $R_{\mu \nu}$ and $R_{55}$ equations as 
well provided that 

\begin{eqnarray}
m^2 =&  - 4 H^2 \qquad & {\rm de \; Sitter} \\
m^2 =&  0       \qquad & {\rm Minkowski} \\
m^2 =&  4 H^2    \qquad & {\rm anti \; de \; Sitter}
\end{eqnarray}

From this it is clear that the radion has negative mass$^2$ for 
de Sitter branes, zero mass for Minkowski branes and positive mass$^2$ for
anti de Sitter branes, the mass$^2$ being proportional to the brane
curvature. This implies that the system of two de Sitter branes is in
fact unstable under small perturbations.

It is straightforward to verify that this solution satisfies the boundary 
conditions at the branes. All components of the metric are continuous 
across the branes. To check the jump conditions on the derivatives of
the metric we go to the Gaussian normal coordinate system at each
brane. Under an infinitesimal coordinate transformation
$x^M = {x'}^M + \Sigma^M(x,r)$ the components of the metric transform as

\begin{equation}
\Delta G_{MN} = G_{TN}\partial_M \Sigma^T + G_{MT}\partial_N \Sigma^T
+ \partial_T G_{MN} \Sigma^T 
\end{equation}

If the condition $\Delta G_{\mu 5} = 0$ is imposed then $\Sigma^5$ is
related to the other $\Sigma^\mu$. Then the transformations of the
remaining components of the metric can be expressed in terms of $\Sigma^5$

\begin{eqnarray}
\Delta G_{55} &=& 2 \partial_5 \Sigma^5 \\
\Delta G_{\mu \nu} &=& f'g_{\mu \nu} \Sigma^5 - 2 f \int dx^5 f^{-1}
\nabla_{\mu} \nabla_{\nu} \Sigma^5 
\end{eqnarray}

We first consider the boundary conditions at $r=0$. We therefore choose
$\Sigma^5$ so as to set $G_{55}$ to one while keeping $\Sigma^5 = 0$ at
$r = 0$ as a boundary condition. Then in the Gaussian normal coordinate
system

\begin{equation}
G_{\mu \nu} \rightarrow \left[ f + \left( 1 + f'\int_0^r dr
f^{-1}\right) \psi \right] g_{\mu \nu}
-2f \int dr \left[ f^{-1} \int_0^r dr f^{-1} \right] \nabla_{\mu}
\nabla_{\nu} \psi
\end{equation}

In this coordinate system the jump condition that has to be satisfied can
be written in the form

\begin{equation}
\Delta \left( \frac{G_{\mu \nu}}{f}\right)' = 0
\end{equation}

It is clear that this condition is satisfied at $r=0$. To verify that the 
matching is satisfied at $r=a$ we must go through the identical procedure
but this time setting $\Sigma^5 = 0$ at $r=a$. Everything else goes
through just as for $r=0$. Hence we conclude that this solution does
indeed satisfy all the boundary conditions.   

Having computed the radion wave function we are now in a position to
determine the couplings of the radion to matter on the two branes. The
interaction Lagrangian is proportional to $ \delta G_{\mu \nu}
\Delta T^{\mu
\nu}$, where

\begin{equation}
\delta G_{\mu \nu} =  \psi (x) g_{\mu \nu}
\end{equation}
and $\Delta T^{\mu \nu}$ is the stress tensor for matter on the brane.
From this it is clear that the radion couples to the trace of the energy
momentum tensor, as expected. The effective coupling constant to matter at
each brane will differ from that of the zero mode graviton which has a
completely different profile in the extra dimension

\begin{equation}
\delta G_{\mu \nu} =  f h_{\mu \nu} (x)
\end{equation}
where $h_{\mu \nu} (x)$ satisfies

\begin{equation}
g^{\alpha \beta}\nabla_{\alpha} \nabla_{\beta} h_{\mu \nu} - 2 H^2 h_{\mu
\nu}
=0
\end{equation} 

If we write the interaction
Lagrangian of the radion with matter on each brane as
$L_{int} = \bar{g} {\psi} \Delta {T^{\mu}}_{\mu}$ then $\bar{g}$ is 
proportional to $f^{-1}$ and hence the radion couples more strongly
to the brane where the warp factor is smaller. This result which
has already been established for flat branes \cite{GT},
\cite{CGR} therefore holds even if the branes are bent.

\section{Effective Theory of the Radion}

In order to investigate the physics of the radion it is of interest to
consider the four dimensional effective theory of the radion valid at
wavelengths long compared to the typical inverse Kaluza Klein masses but
short compared to the inverse of the radion mass. The complete low energy
spectrum will consist of the massless 4D graviton, the radion, and any
Kaluza Klein states that are light relative to the length scales being
probed. We will obtain the effective theory for the radion by substituting
the form of the fluctuation (\ref{eq:solution}) into the linearized
Lagrangian below, and performing the integration over the extra dimension.
We are considering the case where all matter is on the visible brane at
$r=0$.

\begin{equation}
 \int d^4 x d{r} \sqrt{-g} f^2 
\left[\left(2M^3 \right)\frac{1}{2}h^{MN}\delta\left(
R_{MN} - \frac{1}{2} 
 G_{MN} R - 8 \pi G_5 T_{MN} \right) - \frac{1}{2} h^{\mu \nu} \Delta
T_{\mu \nu} \delta(r) \right] 
\end{equation} 
Here $G_5$ is the higher dimensional Newton's constant, and $T_{MN}$ is
the stress tensor from the bulk and brane tensions.

The result has the form

\begin{equation} \int d^4x \sqrt{-g} \left( 2 (2M^3) \int_0^a dr f^{-1}
\right) \left[ \frac{3}{2} \psi \left( g^{\mu \nu} \nabla_{\mu}
\nabla_{\nu} \psi - m^2 \psi \right)\right] + \frac{1}{2} \psi g^{\mu \nu}
\Delta T_{\mu \nu} \end{equation} where we have normalized $f$ to 1 on the
visible brane at $r=0$.  This expression can be generalized to the de
Sitter, flat and anti de Sitter cases by setting $m^2 \rightarrow -4H^2 $,
$m^2 \rightarrow 0 $ and $m^2 \rightarrow 4H^2 $ respectively. We see that
the couplings of the radion to matter on the visible brane are controlled
by the integral in the coefficient of the kinetic term $\int_0^a dr
f^{-1}$. This integral determines the normalization of the radion wave
function. If this integral is big the radion tends to decouple from matter
on the visible brane. Note that the kinetic term of the radion is positive
independently of the brane curvature and so it is not a ghost. We now
consider the implications of this result for physical situations of
interest.

We first consider the case when the branes are Minkowski and we are
localized to the Planck brane which is located at $r=0$. In the limit of
large brane spacing we expect this to smoothly go over to RS2 \cite{RS2},
which does not have a radion. Now $M$ is of order the four dimensional
Planck scale, $m^2 = 0$ and $f = exp(-\alpha r)$. Clearly the integral
$\int_0^a dr f^{-1}$ is dominated by contributions from the region of the
hidden brane at $r=a$ and as it is moved further and further away the
integral grows so that the effects of the radion on the visible brane
disappear in the RS2 limit, as expected.

Now we consider the situation in RS1 \cite{RS1}. This time we are
localized to the
TeV brane which is at $r=0$. Since we are normalizing $f$ to 1 at the
position of the {\bf visible} brane $f = exp(+\alpha r)$ with $\alpha$ and
$M$ of order TeV, and the Planck scale given by ${M_4}^2 \approx 
M^3\int_0^a dr
f$.  Now the integral $\int_0^a dr f^{-1}$ which controls the radion
kinetic term is finite and dominated by contributions from the region of
the visible brane with the result that the radion couples with order TeV
strength to matter on the visible brane. This is in agreement with the
known result \cite{Goldberger}, \cite{Csaki:2000mp}.

Since the mass of the radion in the effective theory is negative for 
de Sitter branes, this configuration is unstable.

Finally we consider the case of anti de Sitter branes. We focus on the
configuration with the warp factor falling as one moves away from the
visible brane at $r=0$. Then $M$ is of order the four dimensional Planck
scale, $m^2 = 4H^2$ and $f = 4 \frac{H^2}{\alpha^2} {\rm cosh^2}
\left[\frac{1}{2}\alpha(r-r_0)\right]$ with $r_0 > 0$. The constraint that
$f=1$ at the position of the visible brane at $r=0$ determines $r_0$ in
terms of $H^2$ and $\alpha$. The Kaluza Klein spectrum of the transverse
traceless modes for this theory is discrete even for infinite brane 
spacing \cite{KR}, with the mass splittings comparable to $H^2$. 

The warp factor has a minimum or `bounce' at $r=r_0$ but does not go to
zero there. It then grows without bound.This has the consequence that the
normalization integral $\int_0^a dr f^{-1}$ is bounded even when $a$ is
made arbitrarily large. If the hidden brane is located behind the bounce
the integral is large and dominated by the region around the bounce and is
of order $\frac{\alpha}{H^2}$. This shows that although the radion
coupling to matter on the visible brane is weak (order $\frac{H}{M^2}$) it
does not vanish even at arbitrarily large separations. Hence this case is
different from the analogous situation with Minkowski branes analysed
above.

In this limit the radion has a mass which is comparable to the masses of
the lightest few (discretely spaced) Kaluza Klein states. Hence these (and
all the KK modes light relative to the scale being probed) must be kept in
the low energy effective theory along with the radion and graviton in
order to determine the complete gravitational physics. In the limit that
$H^2$ is decreased to zero from below the integral once again diverges so
that the RS2 limit is smoothly reached, as we expect.
\\
\\
{\bf Note Added}\\
While this work was being completed we received \cite{BDL} which contains
similar conclusions.
\\
\\
{\bf Acknowledgements} \\
This work was partially supported by the DOE
under contract DE-FGO3-96-ER40956. We would like to thank Emmanuel Katz,
Markus A. Luty, Ann E. Nelson and Lisa Randall for useful conversations
at various stages of this work.


\end{document}